\documentclass[aps,
            prb,
            reprint,
            superscriptaddress,
            nofootinbib]{revtex4-2}
\usepackage{preamble}
\UseRawInputEncoding
\begin{document}

\title{Fidelity Analysis of Adiabatically Driven Donor Spins \\ as Two-Qubit and Ququart Systems}

\author{Brian Michon\,\orcidlink{0009-0001-8178-4788}}
\email{brian.michon@imec.be}
\affiliation{Imec, Kapeldreef 75, 3001 Heverlee, Belgium}
\affiliation{Department of Electrical Engineering, KU Leuven, Kasteelpark Arenberg 10, 3001 Heverlee, Belgium}

\author{James Keppens\,\orcidlink{0000-0001-5698-9549}}
\affiliation{Imec, Kapeldreef 75, 3001 Heverlee, Belgium}
\affiliation{Department of Electrical Engineering, KU Leuven, Kasteelpark Arenberg 10, 3001 Heverlee, Belgium}

\author{Ashutosh Kinikar\,\orcidlink{0000-0002-3085-1274}}
\affiliation{Imec, Kapeldreef 75, 3001 Heverlee, Belgium}
\affiliation{Instituut voor Theoretische Fysica, KU Leuven, Celestijnenlaan 200D, 3001 Leuven, Belgium}

\author{George Simion\,\orcidlink{0000-0002-6880-6161}}
\affiliation{Imec, Kapeldreef 75, 3001 Heverlee, Belgium}

\author{Kristof Moors\,\orcidlink{0000-0002-8682-5286}}
\affiliation{Imec, Kapeldreef 75, 3001 Heverlee, Belgium}

\author{Bart Sor\'ee\,\orcidlink{0000-0002-4157-1956}}
\email{bart.soree@imec.be}
\affiliation{Imec, Kapeldreef 75, 3001 Heverlee, Belgium}
\affiliation{Department of Electrical Engineering, KU Leuven, Kasteelpark Arenberg 10, 3001 Heverlee, Belgium}
\affiliation{Department of Physics, University of Antwerp, Groenenborgerlaan 171, 2020 Antwerp, Belgium}

\date{\today}
 
\begin{abstract}
Donor spin systems host a native Hilbert space whose dimension exceeds that of a qubit, meaning they can be used as qudits. Here we study a \ce{Si{:}P} donor spin system through leakage-aware randomized benchmarking (RB) of native ququart $\mathcal{C}_4$ and encoded two-qubit $\mathcal{C}_2^{\otimes 2}$ Clifford groups. We implement adiabatic ramps to operate electron dipole spin resonance (EDSR) pulses at the ionization point, where the electron is shared halfway between the donor and the interface, and to operate electron spin resonance (ESR) pulses near the interface, motivated by the sensitivity of the effective magnetic field to charge noise at the ionization point. By placing the electron near the ionization point only during EDSR control and using sufficiently long displacement ramp durations, leakage outside the computational basis is strongly suppressed, which is crucial for optimized qudit control. We find in our analysis based on leakage RB that $\mathcal{C}_4$ consistently achieves $\sim 40$--$50\%$ lower (lower-bound) error rates $\varepsilon^{\mathrm{LB}}_\mathrm{PT}$ with respect to $\mathcal{C}_2^{\otimes 2}$, due to its reduced circuit complexity. These results indicate that donor spin qudits benefit from genuine qudit operation as opposed to imposed encoded qubit operation.
\end{abstract}
 
\maketitle

\section{Introduction}
\begin{figure*}
    \centering
    \includegraphics[width=0.95\linewidth]{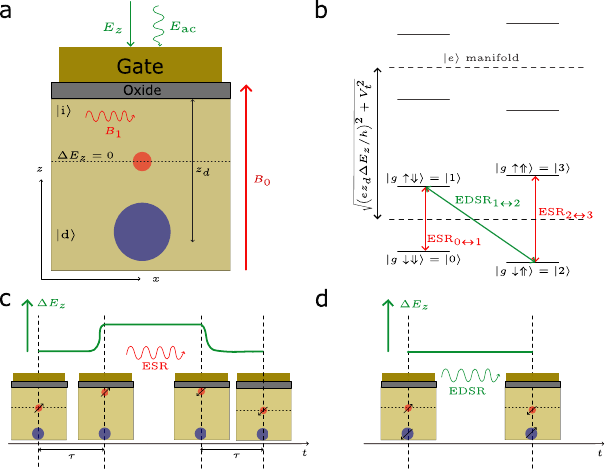}
    \caption{Donor spin device schematic and operation. (a) Schematic visualization of the $\mathrm{Si{:}P}$ donor spin system. The excess electron is drawn in red whereas the phosphorus donor nucleus is shown in blue (not to scale). A straight red arrow indicates the Zeeman field $\mathbf{B}_0$. The implantation depth of the phosphorus atom is denoted by $z_d$ and dotted lines indicate the ionization point, which is the average position of the electron when $\Delta E_z = 0$. (b) Energy level landscape for the donor spin system. Indicated in black are the computational basis states approximated by product states of the degrees of freedom and carrying the computational basis-state labels. Colored arrows indicate ESR (red) and EDSR (green, simplified representation of a second-order process) transitions. As seen from the graph, these transitions alone are sufficient to make a connected graph out of all computational basis states. The faint gray levels are excited orbital states separated from the ground states by an orbital splitting $[(ez_\mathrm{d}\Delta E_z/h)^2 + V_\mathrm{t}^2]^{1/2}$. Transitions between the ground and excited orbital manifolds are not shown. (c) Visualization of ESR drive, in which the electron is displaced near the interface by varying $\Delta E_z$ while applying the ESR pulse. Spin states are indicated by arrows. (d) Visualization of EDSR drive. Spin states are indicated by arrows.}
    \label{fig: DeviceAndLevels}
\end{figure*}

In nearly all quantum computing platforms, the fundamental unit of information is a qubit, which is a two-level quantum system. This paradigm is widely understood, and much effort has been devoted to creating suitable fault-tolerant quantum computing platforms using qubits \cite{nielsen00,shorQEC,FowlerSurfaceCode,moncy2026surfacecodethresholdsqubitfootprints,lu2026quantummagicearlyftqc}. However, most of these qubit systems are embedded within a Hilbert space of larger dimensionality, raising a question about the feasibility of exploiting this extended dimensionality for quantum information processing purposes \cite{campbell_author_2018,Gottesman,koviri2026quantumopticalsynthesishighdimensional}. Studying qubits embedded within qudit Hilbert spaces has led to demonstrations of increased qubit lifetimes \cite{miyahara_decoherence_2023}, efficient implementations of qubit entangling gates through single qudit operations \cite{nikolaeva_universal_2024}, increased readout fidelities \cite{osika_shelving_2022,kehrer_improving_2024} and more exotic applications such as discrete time crystals \cite{goss2026qutrittimecrystalstabilized}.\\ \\
Beyond improving our understanding of the qubit, recent work demonstrates that higher-dimensional systems can be controlled with comparable precision to qubits. In particular, $SU(d)$ control has been experimentally realized for transmons of dimension up to four $d = 4$ \cite{liu_performing_2023}, establishing that superconducting quantum systems may naturally extend to the qubit regime. Similarly, both numerical \cite{low_control_2025} and experimental \cite{ringbauer_universal_2022} studies show that trapped ions provide a compelling platform for qudit-based quantum processing units (QPUs).\\ \\
At the algorithmic level, implementations such as qudit Grover search \cite{shi_efficient_2026} highlight the potential for genuine computational advantages in higher-dimensional systems. These developments motivate further work on qudit-specific error correction schemes and dedicated simulation tools \cite{kabir2026sdimquditstabilizersimulator}. Notably, recent results indicate that error-correction thresholds for qutrits $d = 3$ and ququints $d = 5$ can be comparable to those of qubit-based architectures \cite{keppens_qudit_2025}, reinforcing the viability of the qudit approach. A figure of merit to denote the quality of a gate set implemented on both qubits and qudits is the average gate fidelity. Randomized benchmarking (RB) is a widely-used method to obtain average gate fidelities \cite{Wallman_2014}. This is done by inferring an effective depolarizing error channel in a way that is robust to state preparation and measurement (SPAM) errors. For transmons, it has been shown through RB that native ququart $d = 4$ gates are a viable alternative to gates implemented through the perspective of two encoded qubits using the same basis states \cite{seifert_exploring_2023}. This is a clear indicator that the use of qudits as computational units of information is an interesting avenue to explore future fault-tolerant QPU implementations.\\ \\
This work presents a case study of the viability of operating the \ce{Si{:}P} semiconductor spin system, in which a phosphorus donor atom is implanted in a silicon lattice, as a genuine ququart (Fig.~\ref{fig: DeviceAndLevels}). We drive the system with electron dipole spin resonance (EDSR) and electron spin resonance (ESR) pulses, of which the latter is sensitive to charge noise at the so-called ionization point where the electron is shared halfway between the donor site and the interface. To deal with this sensitivity, we make use of adiabatic ramps to displace the electron such that the electron is only at the ionization point during EDSR transitions, and otherwise sits near the interface. We show that the ququart-native implementation reduces control overhead relative to an encoded two-qubit realization in the same four-level manifold. This leads to improved population-transfer fidelity under charge-noise-induced leakage.\\ \\
This paper is organized as follows. In Sec.~\ref{section: DonorSpin} we discuss the physical system and its properties. In Sec.~\ref{section: Operation} we define all transitions and the driving mechanisms (~\ref{section: Operation}A) and the noise implementation (~\ref{section: Operation}B). Next, we define two gate sets in Sec.~\ref{section: QuditGates}, which are the subject of leakage-aware RB in Sec.~\ref{section: Fidelities}. Finally, we conclude our findings in Sec.~\ref{section: conclusion}.
\section{Donor spin qudits}\label{section: DonorSpin}
Donor spin systems are quantum devices in which a donor atom sits in a silicon lattice and carries an excess electron. If we consider the nuclear spin together with the electron spin of such a system, we can formally identify $(2s+1)(2j+1)$ spin product states with $s = 1/2$ the spin number of the excess electron and $j$ that of the nucleus. In our case, $j=1/2$. Such multilevel systems have been utilized as qubits via several encoding schemes, including the nuclear spin, the combined nuclear-electronic spin, and cat states \cite{entangledNuclei,savytskyy_electrically_2023,yu_schrodinger_2025}. Furthermore, by considering both the electronic and the nuclear spin, donor spin systems with up to $16$ computational basis states have been demonstrated experimentally \cite{fernandez_de_fuentes_navigating_2024}.
The simplest example of a donor spin qudit is the \ce{Si{:}P}
platform. A detailed visual schematic and description of the system is given in Ref.~\cite{tosi_silicon_2017} and we only recall the most important details here shown in Fig.~\ref{fig: DeviceAndLevels}a. We denote $\ket{\mathrm{i}}$ as a state in which the electron is localized at the interface, whereas $\ket{\mathrm{d}}$ denotes a state in which the electron is localized at the donor site.
Following Ref.~\cite{tosi_silicon_2017}, the total Hamiltonian is given by 
\begin{equation}
    \mathcal{H} = \mathcal{H}_\mathrm{orb} + \mathcal{H}_\mathrm{B} + \mathcal{H}_\mathrm{A},
\end{equation}
with $\mathcal{H}_\mathrm{orb}$ the orbital Hamiltonian, $\mathcal{H}_\mathrm{B}$ the Zeeman Hamiltonian and $\mathcal{H}_\mathrm{A}$ the hyperfine coupling between the electron and the nucleus. The orbital section of the qubit can be described by a two-dimensional charge qubit Hamiltonian
\begin{equation}
    \mathcal{H}_\mathrm{orb} = \frac{V_\mathrm{t}\tau_x-\left[ez_\mathrm{d}\left(E_z-E_z^0\right)/h\right]\tau_z}{2},
\end{equation}
where $\tau_x = \ket{\mathrm{i}}\bra{\mathrm{d}} + \ket{\mathrm{d}}\bra{\mathrm{i}}$ and $\tau_z = \ket{\mathrm{i}}\bra{\mathrm{i}}-\ket{\mathrm{d}}\bra{\mathrm{d}}$ are Pauli matrices in the orbital basis $\{\ket{\mathrm{i}}, \ket{\mathrm{d}}\}$. More generally, we will denote Pauli matrices in this charge basis by $\tau_i, \ i\in\{0, x, y, z\}$. $V_\mathrm{t}$ is the orbital coupling term, $e$ the elementary charge, $z_\mathrm{d}$ is the donor implantation depth, $E_z$ the electric field applied at the gate and $E^0_z$ the ionization field for which the electron is shared halfway between the donor and the interface. From now on, we write $\Delta E_z \equiv E_z-E_z^0$ to denote the detuning field. Due to our choice of basis used to construct the Pauli operators, a positive $\Delta E_z$ pulls the electron towards the interface, whereas a negative value pushes it to the nucleus. Here, we assume the region near the interface to be identified with a positive detuning field of $6.5 \ \mathrm{kV/m}$. The terms for Zeeman splitting $\mathcal{H}_\mathrm{B}$ and hyperfine coupling $\mathcal{H}_\mathrm{A}$ are defined as
\begin{equation}\label{eq: Hamiltonian}
    \begin{split}
        \mathcal{H}_\mathrm{B} &= \gamma_eB_0\left[\tau_0 + \left(\frac{\tau_0+\tau_z}{2}\right)\Delta_{\gamma_e}\right]S_z-\gamma_nB_0I_z,\\
        \mathcal{H}_\mathrm{A} &= A\left(\frac{\tau_0-\tau_z}{2}\right)\mathbf{S}\cdot \mathbf{I}.
    \end{split}
\end{equation}
Here, $B_0$ denotes the static magnetic field, $A$ is the hyperfine coupling constant, and $\gamma_e$ and $\gamma_n$ are the electron and nuclear gyromagnetic ratios, respectively. The gyromagnetic ratios are given by $\gamma_e \approx 27.97\,\mathrm{GHz/T}$ and $\gamma_n \approx 17.23\,\mathrm{MHz/T}$. For the simulations presented in this work, we assume $B_0 = 0.4\,\mathrm{T}$ and $A = 117.53\,\mathrm{MHz}$. Moreover, $\Delta_{\gamma_e}$ represents the relative change in gyromagnetic ratio for the electron between a quasi-free electron at the interface and an electron confined to the phosphorus donor, which can be up to $\approx 0.7\%$ \cite{ChangeInDeltaGamma,tosi_silicon_2017}. $S$ and $I$ are spin operators for the electron and nucleus, respectively. Since both are associated with a distinct two-dimensional Hilbert space, the full qudit Hamiltonian including the charge degree of freedom has dimension $8$. The level structure of the system is shown in Fig.~\ref{fig: DeviceAndLevels}b. Within this $8$-dimensional Hilbert space, we choose the orbital ground state spin configurations to be the computational eigenstates. To good approximation, they overlap with product states of the form $\ket{\psi} \approx \ket{g}\ket{m_S}\ket{m_I}$, with $m_S,\ m_I = \pm 1/2$ given a sufficiently large magnetic field $B_0$. Here, $\ket{\mathrm{g}}$ denotes the ground state of the charge degree of freedom, $m_S$ denotes the electronic spin and $m_I$ symbolizes the nuclear spin projection along the $z$-axis. In this work, we use the electron at two operating points; the ionization point ($\Delta E_z = 0 \ \mathrm{kV/m}$) where $\ket{\mathrm{g}} = \left(1/\sqrt{2}\right)\times\left(\ket{\mathrm{d}}+\ket{\mathrm{i}}\right)$  and the interface point ($\Delta E_z = 6.5 \ \mathrm{kV/m}$) where $\ket{\mathrm{g}} = \ket{\mathrm{i}}$. The qudit computational basis states can approximately be mapped onto product states as:
\begin{equation}
\begin{split}
    &\ket{0} \leftrightarrow \ket{\mathrm{g}\downarrow\Downarrow}, \quad \ket{1}\leftrightarrow\ket{\mathrm{g}\uparrow\Downarrow},\\ &\ket{2}\leftrightarrow\ket{\mathrm{g}\downarrow\Uparrow}, \quad \ket{3}\leftrightarrow\ket{\mathrm{g}\uparrow\Uparrow}.
\end{split}
\end{equation}
Using this notation, a single arrow represents the electronic spin, whereas a double arrow represents the nuclear spin. The encoding is chosen according to typical conventions as encountered in \cite{fernandez_de_fuentes_navigating_2024}.

\subsection{Noise implementation}
We model charge noise arising from a single two-level fluctuator (TLF), polarized along the $z$ axis of the device. The instantaneous state of the TLF modifies the total electric field experienced by the qudit system, thereby affecting its dynamics.
The impact of this noise depends strongly on the operating point through the coupling between the orbital degree of freedom described by $\tau_z$ and the donor spin system. At the ionization point, the projected detuning-noise operator is dominated by an off-diagonal coupling in the flip-flop subspace $\{\ket{1}, \ket{2}\}$, producing transverse logical mixing errors \cite{tosi_silicon_2017}. Near the interface operating point, the same perturbation is almost common-mode in the computational subspace; after removing this global phase, the remaining coupling is much weaker and primarily appears as small differential frequency shifts, giving predominantly dephasing-like errors. \cite{cai_quantum_2020}, as reflected in Eq.~\eqref{eq: Hamiltonian}.
We implement TLF noise in a quasi-classical manner using a random telegraph noise (RTN) process. This is characterized by the amplitudes and average lifetimes of the two TLF states. Here, we restrict ourselves to a symmetric TLF model \cite{OneTLF}, where a single noise trace amplitude and lifetime sampled from a Poisson distribution fully describe the noise. The resulting detuning can then be written as an RTN process in continuous time:
\begin{equation}
    \Delta E(t) = \Delta E^0 + A_\mathrm{TLF} s(t), \qquad s(t) \in \{-1, +1\},
\end{equation}
where $\Delta E^0$ denotes the unperturbed detuning field and $A_\mathrm{TLF}$ the amplitude of the RTN trace. In what follows, we use $f_\mathrm{RTN}$ to denote the inverse lifetime of the TLF. 

\section{Operation}\label{section: Operation}
\subsection{Driving mechanisms}
ESR is used to drive transitions between the states $\ket{0}\leftrightarrow\ket{1}$ and $\ket{2}\leftrightarrow \ket{3}$ by applying a magnetic field oscillating at frequency $\omega$ resonant with the corresponding transition frequencies. The magnetic field is oriented along the $x$-axis, $\mathbf{B}_1 = (B_1,0,0)$, resulting in Rabi frequencies on the order of $\gamma_e B_1$ for both transitions (see Figs.~\ref{fig: DeviceAndLevels}b and c).\\ \\
In principle, nuclear magnetic resonance (NMR) could be used to access transitions involving the nuclear spin, thereby enabling full control over the four-dimensional computational basis. However, due to the small nuclear gyromagnetic ratio $\gamma_n$, such transitions are prohibitively slow and are therefore not considered here.\\ \\
Instead, we employ a second-order EDSR mechanism to access the remaining transitions, as is commonly used in flopping-mode qubits \cite{FloppingModeQubit, kinikar_microscopic_2026}. This process arises from the spin-orbit coupling present in the hyperfine interaction (Eq.~\eqref{eq: Hamiltonian}) \cite{tosi_silicon_2017}. The system is driven by an oscillating electric field along the $z$-axis, $\mathbf{E}_\mathrm{ac}(t) = (0,0,E_\mathrm{ac})\cos(\omega t + \varphi)$, which couples to the orbital degree of freedom of the electron (see Figs.~\ref{fig: DeviceAndLevels}b and d). This interaction mediates spin transitions via a flip-flop mechanism, simultaneously flipping the electron and nuclear spins when they are antiparallel.\\ \\
The efficiency of both ESR and EDSR processes depends strongly on the electron's position along the $z$-axis. For EDSR, the transition rate is maximized near the ionization point, where the electron is delocalized between the donor and interface. At this operating point, the hyperfine interaction becomes highly sensitive to the electric field, as quantified by the derivative $\partial A/\partial \Delta E_z$, which is maximized at $\Delta E_z = 0$ \cite{tosi_silicon_2017}. This enhances the effective magnetic driving underlying the EDSR process.\\ \\
In contrast, ESR transitions are adversely affected at the ionization point due to hybridization between spin and charge degrees of freedom. In this regime, magnetic driving can induce unwanted charge transitions, leading to leakage outside the computational subspace and reduced gate fidelity. Similarly, charge noise induces fluctuations of the electron's position which in turn alters the effective magnetic field experienced by the electron. To suppress such effects, ESR operations should be performed when the electron is localized where $\partial A/\partial \Delta E_z$ is minimized. In this work, we choose to localize the electron near the interface, effectively suppressing the hyperfine interaction and bringing the ESR transition frequencies closer together.\\ \\
With this choice of operating regimes, control of the ququart is achieved using two types of pulses: ESR drives for pure electron spin transitions and EDSR drives for flip-flop transitions. The full set of transitions employed is shown in Fig.~\ref{fig: DeviceAndLevels}b. Note that when the electron is localized at the interface the ESR transition frequencies converge to the same value such that $\mathrm{ESR}_{0\leftrightarrow1} =  \mathrm{ESR}_{2\leftrightarrow3}$. At our chosen detuning field, however, the spectral separation is about $6.19 \ \mathrm{MHz}$, allowing for the selectivity necessary to drive the transitions independently.\\ \\
For the driving field strengths, we consider modest amplitudes, with an oscillating electric field of $10\,\text{V/m}$ and a magnetic field amplitude of $100\,\text{\textmu T}$. While these values are conservative \cite{savytskyy_electrically_2023,tosi_silicon_2017}, they are sufficient for a comparative numerical study and have the added effect of increasing gate durations, thereby making the influence of noise more pronounced.

\subsection{Adiabatic ramps between operating points}
\begin{figure*}
    \includegraphics[width=0.85\textwidth]{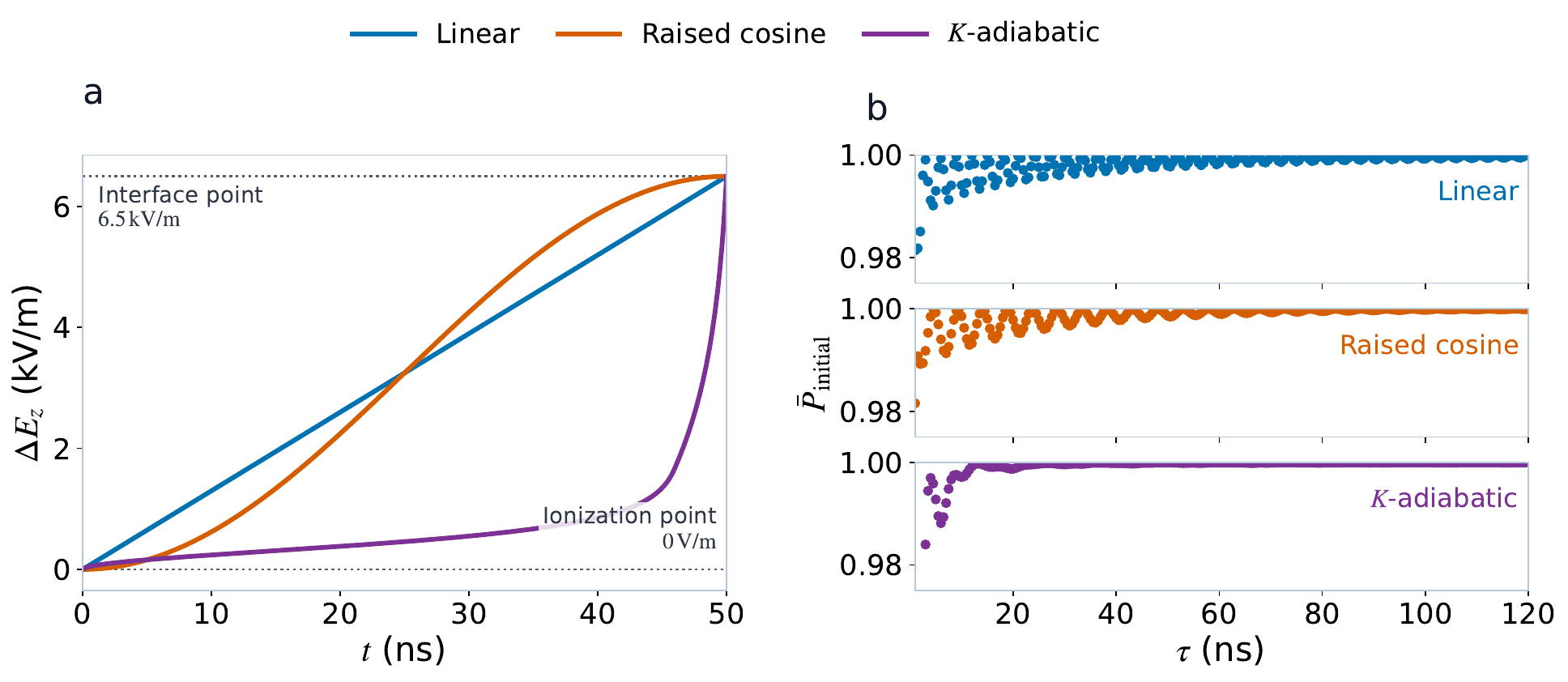}
    \caption{(a) Detuning field profile as a function of time for all ramp types in which the total ramp duration is kept fixed at $\tau = 50 \ \mathrm{ns}$. (b) Initial state survival probability as a function of total ramp duration $\tau$ for all ramp types after averaging over all states after a round trip displacement. A frame update is applied before measurement in order to cancel accumulated phase.}
    \label{fig: CombinedRampPlot}
\end{figure*}

In order to displace the electron to and from the interface, we make use of adiabatic detuning voltage ramps. These are electric fields we apply to the system, shifting the location of the electron, as already demonstrated in Refs.~\cite{tosi_silicon_2017} and \cite{ferraro2021universalsetquantumgates}. In addition to the ramp-time duration, characterized by $\tau$, the shape of the ramp also influences the leakage outside of the charge ground state manifold. In this work we consider three ramp types: a linear ramp, a raised cosine ramp and an adiabatic ramp whose ramp speed is informed explicitly by the avoided crossings. All ramp types can be written as
\begin{equation}
    \Delta E_z(t) = \Delta E_{z,\mathrm{0}} + \left(\Delta E_{z,\mathrm{1}}-\Delta E_{z,\mathrm{0}}\right)s(\alpha),
\end{equation}
where the subscripts $0$ and $1$ stand for the initial and final detuning, respectively. The function $s(\alpha)$ with $\alpha \equiv t/\tau$ describes the ramp profile and depends on the ramp type used.
For the linear ramp, we use
\begin{equation}
    s(\alpha) = \alpha,
\end{equation}
which is the simplest ramp type to implement. The ramp has a constant slope, however, meaning that the ramp does not slow down near avoided crossings where the energy difference between two states is small. This potentially gives rise to substantial leakage during a ramp.
For the raised cosine ramp, we use
\begin{equation}
    s(\alpha) = \frac{1-\cos(\pi \alpha)}{2}.
\end{equation}
This ramp shape has the advantage of having a vanishing ramp speed at the operating points.\\ \\
The final ramp, here named the K-adiabatic ramp, follows the constant adiabatic factor construction used in Ref.~\cite{tosi_silicon_2017}. The relevant energy spectrum contains two regions where nonadiabatic leakage is most likely. The first is the charge avoided crossing, where the donor- and interface-localized charge states would cross in the absence of $V_\mathrm{t}$. The second is the spin-charge avoided crossing caused mainly by the hyperfine interaction term in Eq.~\eqref{eq: Hamiltonian}. Adiabaticity is hardest to maintain whenever the avoided crossing becomes small. Therefore, this ramp type dynamically slows down near the tightest constraint imposed by either avoided crossing. We evaluate the inverse ramp speed required by both avoided crossings $\left(dt/d\Delta E_z\right)_\mathrm{c}$ and $\left(dt/d\Delta E_z\right)_\mathrm{so}$, and choose the larger of the two:
\begin{equation}
    \frac{dt}{d\Delta E_z}
    \propto
    \max\left[
    \left(\frac{dt}{d \Delta E_z}\right)_{\mathrm{c}},
    \left(\frac{dt}{d\Delta E_z}\right)_{\mathrm{so}}
    \right].
\end{equation}
The normalized ramp profile is then obtained by cumulative integration. For $\Delta E_z(s)=\Delta E_{z,\mathrm{0}}+s(\Delta E_{z,\mathrm{1}}-\Delta E_{z,\mathrm{0}})$,
\begin{equation}\label{eq: KAdiabaticRamp}
    \alpha(s)
    =
    \frac{
    \int_0^s w[\Delta E_z(u)]\,du
    }
    {
    \int_0^1 w[\Delta E_z(u)]\,du
    },
\end{equation}
where
\begin{equation}
    w(\Delta E_z)
    =
    \max\left[
    w_{\mathrm{c}}(\Delta E_z),
    w_{\mathrm{so}}(\Delta E_z)
    \right]
\end{equation}
is the local inverse sweep speed. The ramp used in the simulation (Eq.~\eqref{eq: KAdiabaticRamp}) is the inverse
map $s(\alpha)$, so that equal increments of physical time correspond to equal
increments of accumulated adiabatic weight $\int_0^sw\left[\Delta E_z(u)\right]du$. The ramp shapes are compared for a fixed total duration of $\tau = 50 \, \mathrm{ns}$ in Fig.~\ref{fig: CombinedRampPlot}a.
Most notably, the K-adiabatic ramp used here and in Ref.~\cite{tosi_silicon_2017} bears resemblance to the filter used in Ref.~\cite{ferraro2021universalsetquantumgates}, where two linear ramp profiles with different slopes are stitched together.\\ \\
We consider an experiment in which we prepare the ququart in a specific state and operating point (here chosen to be the ionization point). We displace the electron to the interface and back (without a waiting time) and calculate the survival probability after a frame update correcting the additional phase gained during time evolution. The result is shown in Fig.~\ref{fig: CombinedRampPlot}b, where we average over four runs, each associated with the initial state being a computational basis state.\\ \\
In terms of survival probability, all ramp types take the shape of an exponential envelope in which oscillations take place. As is clear from the numerical experiment, the raised cosine filter and especially the adiabatic filter significantly reduce the oscillations in survival probability, which improves the reliability of operating the ququart.

\section{Qudit gate sets}\label{section: QuditGates}
The aim in gate-based quantum computing is to find a finite set of unitary gates acting on the $d$-level system whose compositions can approximate any $U \in SU(d)$ to arbitrary precision \cite{nielsen00,brylinski2001universalquantumgates}. In practice, we implement such gate sets by controlling the quantum system using E(D)SR pulses as illustrated in Figs.~\ref{fig: DeviceAndLevels}c and d. Since our system has a connected coupling graph (see Fig.~\ref{fig: DeviceAndLevels}b), it is possible to implement effective two-level subspace rotations, also known as Givens rotations, between any two distinct computational levels $\ket{j}$ and $\ket{k}$ \cite{ringbauer_universal_2022}:
\begin{equation}
    \mathcal{U}_{jk}(t) = G_{jk}\left(\theta\right) = \exp\left(-\frac{i}{\hbar}\theta(t)\Lambda^{S}_{jk}\right).
\end{equation}
Here, $\Lambda^{S}_{jk}$ represent the symmetric generalized Gell-Mann (GGM) matrices \cite{bertlmann_bloch_2008}:
\begin{equation}
    \Lambda^{S}_{jk} \equiv \ket{j}\bra{k} + \ket{k}\bra{j}, \quad j\neq k.
\end{equation}
In our work, these Givens rotations represent the ESR and EDSR transitions as denoted in Figs. ~\ref{fig: DeviceAndLevels}b,c,d. Note that these rotations are effective two-level rotations within the qudit Hilbert space along the $x$-axis within the rotating wave approximation (RWA), in which fast-rotating terms are neglected. However, it is in principle possible to use any rotation axis in the $xy$-plane (for example $Y$ rotations) by using a nontrivial phase for the driving fields responsible for activating $\Lambda^{S}_{jk}$ rotations.
Using this principle, we are able to implement any $SU(4)$ gate by using a fixed two-level rotation sequence \cite{seifert_exploring_2023}:
\begin{equation}\label{eq: Decomposition}
	\begin{aligned}
		&U_\mathrm{Gate}(\boldsymbol{\phi}_1,...,\boldsymbol{\phi}_4;\theta_1, ...,\theta_6) \\ &\, = Z(\boldsymbol{\phi}_4)\, Y_1(\theta_6)\, Y_2(\theta_5)\, Y_3(\theta_4)\, Z(\boldsymbol{\phi}_3) \\
		&\, \hphantom{=} \times Y_1(\theta_3)\, Y_2(\theta_2)\, Z(\boldsymbol{\phi}_2)\, Y_1(\theta_1)\, Z(\boldsymbol{\phi}_1).
	\end{aligned}
\end{equation}
Here, $Y_i$ stands for a $Y$-rotation in the $\{\ket{i-1}, \ket{i}\}$ subspace, parametrized by an angle $\theta_j$ and $Z\left(\boldsymbol{\phi}_i\right)$ is a phase matrix of the form \cite{seifert_exploring_2023}
\begin{equation}
	Z = \mathrm{diag}\left(1, e^{i\phi_{i,1}}, e^{i\left(\phi_{i,1}+\phi_{i,3}\right)}, e^{i\left(\phi_{i,1}+\phi_{i,2} + \phi_{i,3}\right)}\right),
\end{equation}
where $\phi_{i,n}, \ n\in \{1,2,3\}$ is the $n$th entry of the $\boldsymbol{\phi}_i$ vector, which can be chosen to obtain the desired angle. In this work, all phase gates, i.e. $Z$-gates, are implemented virtually (all phases are absorbed into the driving fields) such that they have a near-perfect fidelity.
\\ \\
An important yet non-universal gate set to help realize such $U$ is the single-qudit Clifford gate set $\mathcal{C}_{d}$ \cite{Gottesman,ringbauer_universal_2022}. Consider the generalized Pauli group denoted by $\mathcal{P}_{d}$. Its generating set is given by
\begin{equation}
    \mathcal{P}_{d} = \braket{\mathbf{X}_{d}, \mathbf{Z}_{d}}.
\end{equation}
Here, $\mathbf{X}_{d}$ and $\mathbf{Z}_{d}$ take the general form \cite{Farinholt_2014,keppens_qudit_2025}
\begin{equation}
    \begin{split}
        \mathbf{X}_{d}^p\ket{k} &= \ket{k\oplus p \ \mathrm{mod} \ d},\\
        \mathbf{Z}_{d}^q\ket{k} &= \omega^{q\cdot k}_{d}\ket{k},
    \end{split}
\end{equation}
with $p,q, k\in\{0, 1, ..., d-1\}$ and  $\omega_{d} \equiv e^{2\pi i / d}$. Boldfaced symbols are used to distinguish these gates from effective qubit rotations as used in the gate decomposition of eq.~\eqref{eq: Decomposition}. The single-qudit Clifford group $\mathcal{C}_{d}$ is the normalizer of the Pauli group, and is defined such that $\mathcal{C}_{d} = \{U \in  \ U(d) \ | \ U\mathcal{P}_{d}U^\dagger = \mathcal{P}_{d}\}$. It is generated by \cite{Farinholt_2014,FTQudits} 
\begin{equation}
    \mathcal{C}_{d} = \braket{F_{d}, S_{d}, Z_{d}}.
\end{equation}
In this expression, $F_{d}$ and $S_{d}$ are defined up to a phase such that \cite{keppens_qudit_2025}
\begin{equation}
    \begin{split}
        F_{d}\ket{k} &= \frac{1}{\sqrt{d}}\sum_{a=0}^{d-1}\omega^{k\cdot a}_{d}\ket{a},\\
        S_{d}\ket{k} &= \omega_{d}^{\left(k-d-2\right)k/2}\ket{k}.
    \end{split}
\end{equation}
There are two groups of interest, discussed below.
\subsection{Native ququart Clifford group $\mathcal{C}_4$}
Let us first consider a group structure based on the natively available Hilbert space of a ququart. Its Pauli group, $\mathcal{P}_4$ in this case, is generated by
\begin{equation}
    \mathcal{P}_4 = \braket{X_4, Z_4}.
\end{equation}
This Pauli group is associated with $\mathcal{C}_4 = \{U \in U(4) |\  U\mathcal{P}_4U^\dagger = \mathcal{P}_4\}$, which in turn is generated by 
\begin{equation}
    \mathcal{C}_4 = \braket{F_4, S_4, Z_4}.
\end{equation}
This single-qudit Clifford group is known to contain 768 unique elements and preserves the Pauli group structure defined by $\mathbf{X}_4$ and $\mathbf{Z}_4$ \cite{seifert_exploring_2023}. In this work, we only consider $F_4$ to be a physical generator, as phase operations are effected by adjusting the phase of the driving field.

\subsection{Encoded two-qubit Clifford group $\mathcal{C}_2^{\otimes 2}$}
\begin{figure}[tb]
    \centering
    \includegraphics[width=\linewidth]{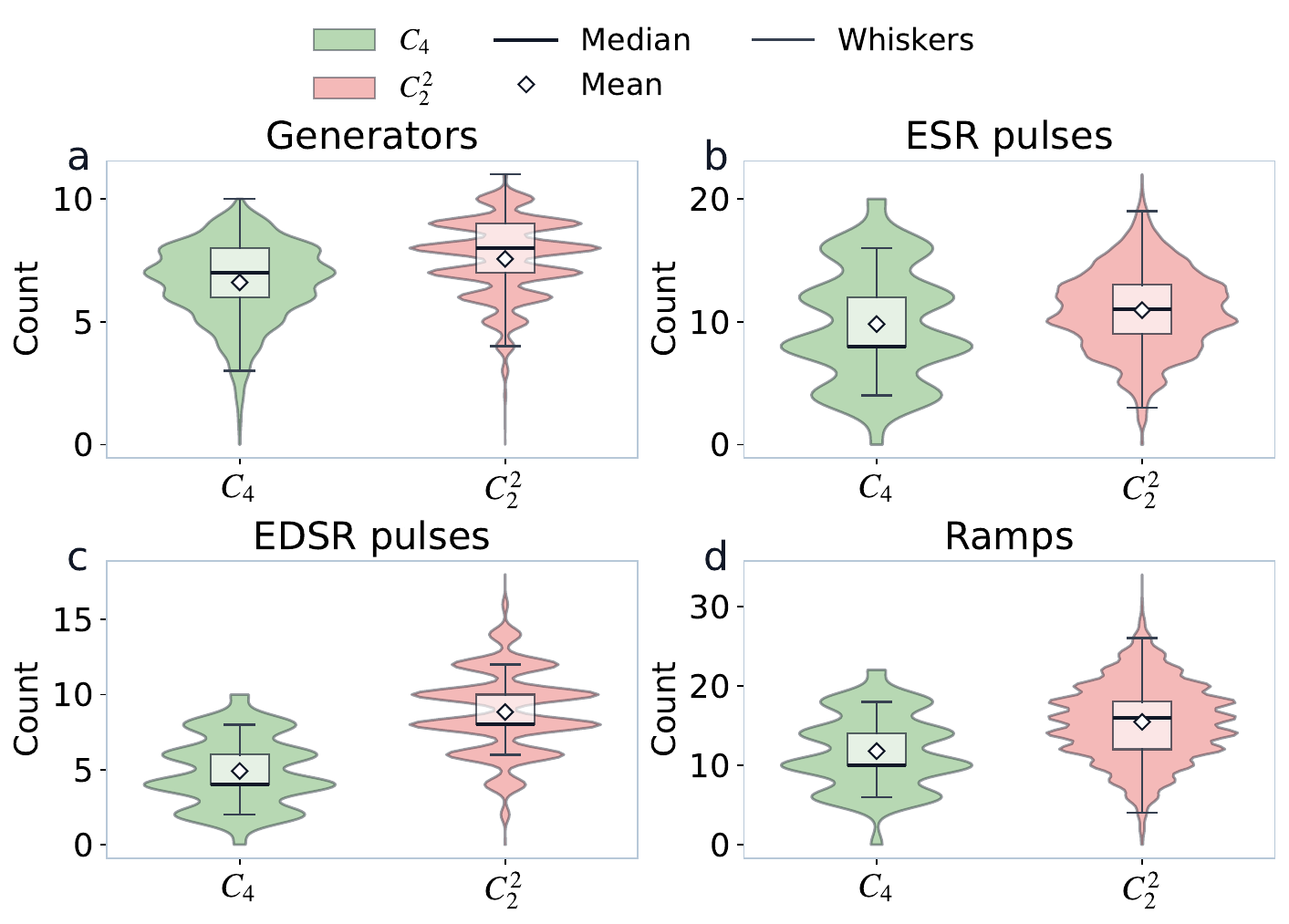}
    \caption{Violin distributions indicating the resource counts for the $\mathcal{C}_4$ and $\mathcal{C}_{2}^{\otimes 2}$ gate sets. (a) Distribution for the number of required generators (width indicates relative frequency). (b) Distribution for the number of used ESR pulses. (c) Distribution for the number of used EDSR pulses. (d) Distribution for the number of used displacement operations.}
    \label{fig: ResourceCounts}
\end{figure}

It is possible to consider a ququart as two encoded qubits, by considering the logical identification \cite{seifert_exploring_2023}
\begin{equation}
    \ket{0} \leftrightarrow \ket{00}, \ \ket{1} \leftrightarrow\ket{01}, \ \ket{2}\leftrightarrow\ket{10}, \ \ket{3}\leftrightarrow\ket{11}.
\end{equation}
In this isomorphism $\mathbb{C}^4\cong\mathbb{C}^2\otimes \mathbb{C}^2$ we can define
\begin{equation}
    \begin{split}
        X_1 &= X\otimes I, \quad  Z_1 = Z\otimes I,\\
        X_2 &= I\otimes X, \quad Z_2 = I\otimes Z.
    \end{split}
\end{equation}
Here, $X$ and $Z$ are the usual Pauli operators for qubits. The Pauli group $\mathcal{P}_2^{\otimes 2}$ is now generated by
\begin{equation}
    \mathcal{P}_2^{\otimes 2} = \braket{X_1, X_2, Z_1, Z_2}.
\end{equation}
Its normalizer, the encoded two-qubit Clifford group $\mathcal{C}_2^{\otimes 2} = \{U\in U(4)| \ U\mathcal{P}_2^{\otimes 2}U^\dagger = \mathcal{P}_{2}^{\otimes 2}\}$ is generated by \cite{nielsen00}
\begin{equation}
    \mathcal{C}_2^{\otimes 2} = \braket{H_1, H_2, S_1, S_2, \mathrm{CNOT}},
\end{equation}
where the generator subscripts indicate the same type of structure as for the Pauli operators. Moreover, $H$ and $S$ represent the conventional Hadamard and $S$-gates for qubits, respectively. Note that $Z_i$ is redundant since $Z_i = S^2_i$ for qubits. Moreover, there is now an additional two-qubit gate $\mathrm{CNOT}\ket{\mathrm{c}, \mathrm{t}} = \ket{\mathrm{c}, \mathrm{t}\oplus \mathrm{c}}$. The two-qubit Clifford group thus preserves the tensor structure and contains $11 \, 520$ unique elements, which are generated using more physical operators than the single-qudit Clifford group \cite{seifert_exploring_2023}.\\ \\
We build both gate sets and implement them explicitly using Eq.~\eqref{eq: Decomposition}. The phase values required to implement each gate are found in Ref.~\cite{seifert_exploring_2023}. In Fig.~\ref{fig: ResourceCounts}, we give the statistical distribution of the resource counts necessary to implement all Clifford elements.
More specifically, we compare the required number of generators (Fig.~\ref{fig: ResourceCounts}a), the number of E(D)SR pulses (Fig.~\ref{fig: ResourceCounts}b-c) and the number of displacement ramps (Fig.~\ref{fig: ResourceCounts}d). As a general trend, we notice that the mean and median for $\mathcal{C}_4$ are lower than for $\mathcal{C}_2^{\otimes 2}$. Additionally, the latter has more outliers for all metrics. The general conclusion we draw is that the implementation of the ququart Clifford elements is more economical in terms of pulse count as compared to their encoded two-qubit counterpart.

\section{Average gate fidelities}\label{section: Fidelities}
We turn to comparing the two different gate sets. First, we discuss how RB is used to obtain gate fidelities. However, standard RB assumes all errors to stay confined to the computational subspace, which is not the case in our system. Therefore, standard RB systematically overestimates the fidelity. Therefore, we opt for leakage-aware RB which is more suited to describe the average gate fidelity when leakage into the excited-state manifold of the setup under consideration here (see Fig.~\ref{fig: DeviceAndLevels}b) is present \cite{chen_randomized_2025}.
\begin{figure*}[!htbp]
\centering
\includegraphics[width=0.95\textwidth]{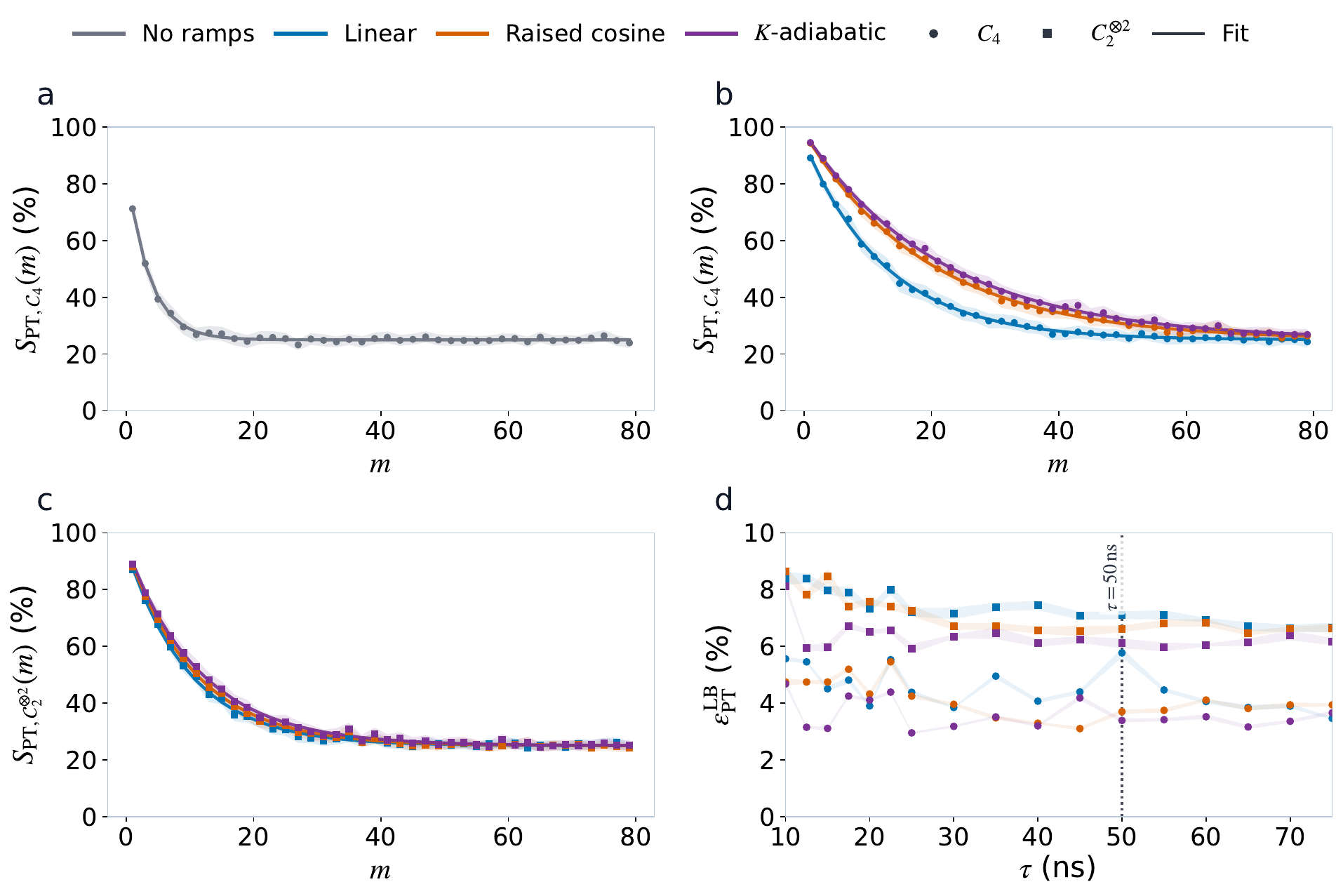}
\caption{Overview of RB results. Numerical metrics for leakage-aware RB are reported in Table~\ref{tab:ramp-performance}. All simulations use $n_\mathrm{seeds}=10$, $n_\mathrm{trials}=150$, $A_\mathrm{TLF}=10\,\mathrm{V/m}$, and $f_\mathrm{RTN}=10\,\mathrm{Hz}$. Shaded regions denote $99\%$ confidence intervals; dots show simulated data and solid lines the fitted decay curves. For ramped RB, the ramp duration is $\tau=50\,\mathrm{ns}$ unless stated otherwise.
(a) Leakage-aware survival probability $S_{\mathrm{PT},4}(m)$ without adiabatic ramps.
(b),(c) Leakage-aware survival probabilities $S_{\mathrm{PT}}(m)$ for ramped RB experiments with adiabatic ramps using (b) ququart ($\mathcal{C}_4$) encoding and (c) two-qubit ($\mathcal{C}_2^{\otimes 2}$) encoding.
(d) Absolute lower-bound error $\varepsilon_\mathrm{PT}^\mathrm{LB}=1-(1+3r_\mathrm{PT})/4$ for $\mathcal{C}_4$ (circles) and $\mathcal{C}_2^{\otimes 2}$ (squares) as a function of ramp duration $\tau \ge 10\,\mathrm{ns}$. The vertical dotted line marks $\tau=50\,\mathrm{ns}$, used in panels (b) and (c).
Leakage-aware survival probabilities in panels (a)–(c) are fitted with $1/d + A_\mathrm{PT} r_\mathrm{PT}^m$.}
\label{fig:RBSummary}
\end{figure*}

\begin{table*}[t]
\caption{\label{tab:ramp-performance}%
Performance metrics for the different ramp types and Clifford groups
at $\tau=50~\mathrm{ns}$. Results without ramps are omitted.}
\begin{ruledtabular}
\begin{tabular}{
    ll
    D{.}{.}{1.4}
    D{.}{.}{2.1}
    D{.}{.}{1.4}
    c
}
\multirow{2}{*}{Ramp type}
&
\multirow{2}{*}{Group}
&
\multicolumn{2}{c}{$S(m)$}
&
\multicolumn{2}{c}{$S_{\mathrm{PT}}(m)$}
\\
&
&
\multicolumn{1}{c}{$p$}
&
\multicolumn{1}{c}{$\overline{F}_{\mathrm{RB}}$ (\%)}
&
\multicolumn{1}{c}{$r_{\mathrm{PT}}$}
&
\multicolumn{1}{c}{$\overline{F}_{\mathrm{PT}}$ (\%)}
\\
\colrule

\multirow{2}{*}{Linear}
&
$\mathcal{C}_4$
&
0.9274
&
94.6
&
0.9247
&
92.5--94.4
\\
&
$\mathcal{C}_2^{\otimes 2}$
&
0.9107
&
93.3
&
0.9040
&
90.4--92.8
\\[2pt]

\multirow{2}{*}{Raised cosine}
&
$\mathcal{C}_4$
&
0.9532
&
96.5
&
0.9504
&
95.0--96.3
\\
&
$\mathcal{C}_2^{\otimes 2}$
&
0.9198
&
94.0
&
0.9108
&
91.1--93.3
\\[2pt]

\multirow{2}{*}{$K$-adiabatic}
&
$\mathcal{C}_4$
&
0.9585
&
96.9
&
0.9551
&
95.5--96.6
\\
&
$\mathcal{C}_2^{\otimes 2}$
&
0.9245
&
94.3
&
0.9172
&
91.7--93.8
\\
\end{tabular}
\end{ruledtabular}
\end{table*}
\subsection{Clifford randomized benchmarking}
In this numerical experiment, the native ququart Clifford group $\mathcal{C}_4$ is compared with the two-qubit Clifford group $\mathcal{C}_{2}^{\otimes 2}$. The protocol samples $m$ Clifford operations from either Clifford group $\mathcal{C}_\mathrm{sample} = \mathcal{C}_m \circ \mathcal{C}_{m-1}\circ ...\circ \ \mathcal{C}_{1}$, which are decomposed into pulses necessary to construct the required unitary, after which the inverse operation $\mathcal{C}_\mathrm{inv}$ is implemented. In order to reduce the computational overhead, pulse propagators are precomputed for the noiseless case and both energetic states of the TLF. Similarly, the Clifford elements of both groups are precomputed. Therefore, within a RB loop, it is only necessary to explicitly evaluate time-evolution operators around switching events. Let $\rho_m$ be the state after implementing the sequence of $m$ Clifford gates followed by the inversion gate and, which ideally leads to the final state being $\ket{\psi_\mathrm{init}} = \ket{0}$. The fidelity for this sample is given by the survival probability 
$S(m) = \braket{\psi_\mathrm{init}(m)|\rho_m|\psi_\mathrm{init}(m)}$,
with $0 \leq S(m) \leq 1$. In order to obtain fair statistics, we define a seed that determines the starting value for the random Clifford sequence generator and the initial TLF configuration. This is useful for ensuring an honest comparison for all ramp types, such that the results use the same random sequences across all ramp types. Each seed has $n_\mathrm{trials}$ trials, such that the total number of samples is $n_\mathrm{seeds}\times n_\mathrm{trials}$. Averaged fidelities for each $m$ can now be fitted to an exponential function of the form 
\begin{equation}
	S(m) = A p^m + B,
\end{equation}
where $p$ is the effective depolarizing parameter, which can be converted into an average gate fidelity $\overline{F}$ for a qudit of dimension $d$ as follows \cite{Magesan_RB}:
\begin{equation}
	\overline{F} = 1 - \frac{d-1}{d}\left(1-p\right).
\end{equation}
Here, the prefactor $\left(d-1\right)/d$ converts the error per Clifford $1-p$ into an average gate infidelity. Hence, the error can be defined as $\varepsilon = \left(d-1\right)\left(1-p\right)/d$.
\subsection{Leakage-aware randomized benchmarking}
Standard CRB fails to accurately describe the average gate fidelity in the presence of leakage \cite{RBLimits, LeakageRB,sutherland2025subspaceleakageerrorrandomized,Wallman_2016}. To take leakage outside of the computational basis into account in our fidelity analysis, we consider the leakage-aware RB procedure of Ref.~\cite{chen_randomized_2025}.
Let $\Pi_\mathrm{comp}$ be the projector onto the computational subspace such that the retained population $R(m)$ is given by
\begin{equation}
    R(m) = \mathrm{Tr}\left[\Pi_\mathrm{comp}\rho_m\right].
\end{equation}
Hence, the leakage probability $L(m)$ is given by
\begin{equation}
    L(m) = 1-R(m) = 1-\mathrm{Tr}\left[\Pi_\mathrm{comp}\rho_m\right].
\end{equation}
Following Ref.~\cite{chen_randomized_2025}, we now take into account the (reversible) leakage outside the computational subspace caused by spin-orbit interactions by defining the averaged measurement survival $S_\mathrm{PT}(m)$ (where the subscript $\mathrm{PT}$ indicates that we consider population transfer outside of the computational basis):
\begin{equation}
    S_\mathrm{PT}(m) = S(m) + \frac{L(m)}{d}.
\end{equation}
The term $L(m)/d$ assigns leaked population the probability of a uniformly
random computational outcome. This removes the need to interpret leakage as a
one-way loss process, since the leakage in our system is reversible. The population transfer decay parameter
$r_\mathrm{PT}$ is extracted from \cite{chen_randomized_2025}
\begin{equation}
    S_\mathrm{PT}(m)
    \approx
    \frac{1}{d}+A_\mathrm{PT} r_\mathrm{PT}^m .
\end{equation}
The resulting decay parameter does not uniquely determine a single average gate
fidelity in the presence of (reversible) leakage. Instead, it bounds the average
fidelity as \cite{chen_randomized_2025}
\begin{equation}
    r_\mathrm{PT}
    \leq
    \overline{F}
    \leq
    1-\frac{d-1}{d}\left(1-r_\mathrm{PT}\right).
\end{equation}
For the ququart computational space used here, $d=4$, whence
\begin{equation}
    r_\mathrm{PT}
    \leq
    \overline{F}
    \leq
    \frac{1+3r_\mathrm{PT}}{4}.
\end{equation}
We report $r_\mathrm{PT}$ together with these fidelity bounds as the primary
leakage-aware benchmarking metric.

In a first experiment, we numerically conduct a leakage-aware RB experiment for the $\mathcal{C}_4$ group implemented on the donor spin ququart in which the electron is parked at the ionization point. Here, we use a noise amplitude of $A_\mathrm{RTN} = 10 \, \textnormal{V/m}$ with a switching frequency $f_\mathrm{RTN} = 10 \, \mathrm{Hz}$. The result is shown in Fig.~\ref{fig:RBSummary}a for $n_\mathrm{seeds} = 10$ and $n_\mathrm{trials} = 150$.
We recover the expected behavior that survival probability is very low when implementing ESR drives at the ionization point, as a consequence of the high sensitivity of the effective magnetic field to charge noise \cite{tosi_silicon_2017}. Note that in our simulations, charge noise is quasistatic with very rare switching events. The mean dwell time is longer than the gate durations and the TLF therefore acts primarily as a detuning offset during a given RB shot. Consequently, the exponential decays should be interpreted as randomized and sequence-averaged effective decay parameters rather than implying Markovian noise.\\ \\
In Figs.~\ref{fig:RBSummary}b and c we repeat the experiment for both the $\mathcal{C}_4$ and $\mathcal{C}_2^{\otimes 2}$ groups, now using the different ramp types, each shaped such that the total ramp duration is $50 \, \mathrm{ns}$ in order to meaningfully compare the effect of the ramp shape on the average gate fidelity. We notice a substantial increase in fidelity bounds, owing to the fact that the electron is now placed in a first order charge noise-insensitive region during ESR operations, which are otherwise severely affected by this noise at the ionization point. More specifically considering $\mathcal{C}_4$, the upper bound on fidelity for the case without ramps is limited to $\overline{F}_\mathrm{PT, no \ ramps}\left(\mathcal{C}_4\right) = 81.7\%$, whereas when including ramps we find an upper bound $\overline{F}_\mathrm{PT, ramps}^{\mathrm{UB}}\left(\mathcal{C}_4\right) = \{94.4\%,96.3\%,96.6\%\}$, depending on ramp type. The remaining errors are mainly due to nonadiabatic errors during ramps and dephasing. Finally, Table~\ref{tab:ramp-performance} summarizes the quantitative results and includes predictions according to standard RB. We notice that the suggested fidelity through standard RB systematically exceeds the upper bound given by leakage-aware RB, consistent with the conclusions of Ref.~\cite{chen_randomized_2025}.
\subsection{Ramp time optimization}
As discussed, ESR transitions are very sensitive to charge noise at the ionization point, because $\partial A / \partial E_z$ is very large, altering the experienced magnetic field and thus disrupting population transfer. To decrease this sensitivity, we opt for a region in which this derivative is small such that the impact of noise on the ESR transitions is maximally suppressed. An example of such a region is the interfacial region where the electron is effectively decoupled from the nucleus. However, fully decoupling the electron from the nucleus would converge the ESR transition frequencies of the $\ket{0}\leftrightarrow\ket{1}$ and $\ket{2}\leftrightarrow\ket{3}$ transitions, destroying the addressability of the device. For that reason, we choose a point near the interface that is at the same time far removed from the ionization point, yet not too localized at the interface. \\ \\
The four ramp types discussed previously aim to displace the electron to a favorable position for ESR transitions, without causing excess leakage into the excited manifold of the orbital subspace. This immediately defines a trade-off: pulses that are completed faster (small $\tau$) reduce the additional gate duration, limiting the excess noise exposure of the qudit, but enhance the nonadiabatic error. Therefore, we conduct a numerical leakage-aware RB experiment in which we allow $\tau$ to vary. In Fig.~\ref{fig:RBSummary}d we compare the lower bound for the error $\varepsilon_\mathrm{PT}^\mathrm{LB} = 1-\left(1+3r_\mathrm{PT}\right)/4$ for $\mathcal{C}_4$ and $\mathcal{C}_2^{\otimes}$ as a function of $\tau$, which is the lower bound gate infidelity inferred from the upper bound for the fidelity through the parameter $r_\mathrm{PT}$. These results are obtained by averaging the individual results of 10 RB seeds of 150 trials each, which effectively gives us a sample size of $10 \times 150 = 1 \, 500$ samples per $\tau$. As seen in the figure, there is a consistent $40$--$50\%$ reduction of the lower bound error for $\mathcal{C}_4$ compared to $\mathcal{C}_2^{\otimes 2}$. The remaining difference in fidelity after the electron is moved towards the interface for ESR transitions is directly related to total noise exposure: $\mathcal{C}_2^{\otimes 2}$ requires not only more ESR pulses, but also more EDSR pulses and more ramps. Leakage during these ramps accumulates as well as additional dephasing compared to $\mathcal{C}_4$. Since the ramps are short with respect to the actual pulse durations, the ramp time seems to have little effect on the resulting error. Note, however, that in systems where the ramp duration becomes comparable to typical pulse durations, one must find a sweet spot between ramping slow enough to avoid nonadiabatic errors, but fast enough to avoid excessive noise exposure.

\section{Conclusion}\label{section: conclusion}
In this paper we studied the operation of a \ce{Si{:}P} donor spin system as both a native ququart and an encoded two-qubit device under realistic charge noise. Using ESR- and EDSR-driven control, we performed a fidelity analysis to compare the robustness and computational efficiency of the corresponding gate sets to identify a systematic discrepancy in the observed performance between the two encodings.\\ \\
Using leakage-aware RB, we quantified the impact of realistic charge noise on \ce{Si{:}P} donor spin ququarts and evaluated strategies for mitigating its effect. Through multiple adiabatic ramp shapes, we demonstrated a significant improvement in fidelity upon displacing the electron towards the interface, except when an EDSR drive is applied (with the electron at the ionization point). Moreover, by comparing the native ququart Clifford group $\mathcal{C}_4$ with an encoded two-qubit Clifford group $\mathcal{C}_2^{\otimes 2}$, we obtained a $40$--$50\%$ lower-bound error reduction in favor of a native qudit implementation under realistic charge noise. These results further motivate the use of donor-based qudits as higher-dimensional computational units. The methodology presented here can be readily extended to other donor species and qudit platforms.\\ \\
Overall fidelities may be increased by replacing the rectangular filter for the drive pulses by suitable envelopes for the quantum system and using stronger driving fields, which falls outside the scope of this numerical analysis. In either case, the structure of the two groups is compared in a statistical analysis that indicates a more economical implementation of the native ququart gate set, such that the discrepancy is structural rather than an artifact of the driving conditions.
%\newpage
\bibliographystyle{apsrev4-2}
\bibliography{ref}
\end{document}